\documentclass[aps,prb,twocolumn,showpacs]{revtex4}

\usepackage{amsmath}
\usepackage{epsfig}
\usepackage{graphics}

\renewcommand{\Re}{{\rm Re}}

\def\ba{\begin{array}}
\def\ea{\end{array}}

\begin{document}


\title{Nonequilibrium stabilization of charge states in double quantum dots}


\author{Udo Hartmann}
\email[]{hartmann@theorie.physik.uni-muenchen.de}
\author{Frank K. Wilhelm}
\affiliation{Physics Department and CeNS, Ludwig-Maximilians-Universit\"at, 
Theresienstr. 37, D-80333 M\"unchen, Germany}


\date{\today}

\begin{abstract}
We analyze the decoherence of charge states in double quantum dots
due to cotunneling.  
The system is treated
using the Bloch-Redfield generalized master equation for the Schrieffer-Wolff
transformed Hamiltonian. 
We show that the decoherence, characterized through 
a relaxation $\tau_{r}$ and a dephasing time $\tau_{\phi}$, can be controlled
through the external voltage and that the optimum point, where these 
times are maximum, is not necessarily in equilibrium. We outline the mechanism
of this nonequilibrium-induced enhancement of lifetime and coherence. 
We discuss the relevance
of our results for recent charge qubit experiments. 
\end{abstract}


\maketitle


The loss of quantum coherence is a central paradigm of modern
physics. It not only governs the transition between the quantum-mechanical
and the classical world, but has recently also gained practical importance in 
the context of engineering quantum computing devices. 
Decoherence naturally
occurs in small quantum systems coupled to macroscopic heat baths. A huge 
class of such baths generates Gaussian noise and can hence be mapped on 
an ensemble of harmonic oscillators as in the spin-boson 
model \cite{Leggett}. This can even apply, if the fundamental degrees
of freedom of the bath are fermions, 
as it is, e.g., the case if the bath
is a linear electrical circuit \cite{EPJ,ZimanyiPRL}, which is 
producing Gaussian Johnson-Nyquist noise. 

In this Rapid Communication, we study a generically different system: a 
double quantum dot coupled to electronic leads. Such systems are studied as
realizations of quantum bits \cite{Blick,UdoFrank,Hayashi}. 
The position (either left or right dot) of an 
additional spin-polarized electron is used as the computational basis of a 
{\em charge} qubit as realized in Ref. \cite{Hayashi}.
For another proposal of a charge qubit in semiconductors,
see Ref.~\cite{Hollenberg}.

Our system simultaneously couples to two
distinct reservoirs of real fermions. Other than
oscillator bath models, this allows for the application of a voltage
between these reservoirs as a new parameter for controlling decoherence. 
The voltage creates nonequilibrium between the baths, which to the best of
our knowledge has not been studied yet in the literature on open
quantum systems.

We study the dynamics of the reduced density matrix and identify
the usual two modes of decoherence, dephasing and relaxation:
Dephasing is the loss of phase information, manifest
as the decay of coherent 
oscillations. This corresponds to the time evolution of the off-diagonal 
elements of the reduced density matrix in the energy basis. 
Relaxation is the process during which a quantum system
exchanges energy with the environment 
and ends up in a stationary state. This
is described through the evolution of the diagonal density matrix 
elements. 
We are going to show that, surprisingly, the charge states can be stabilized
by external nonequilibrium, i.e., the relaxation time is longest at a 
well-defined finite voltage.
We will show, that
this working point is also very favorable in terms of dephasing but 
competes with another local maximum at zero voltage. Our theory 
should also have applications in 
other systems.

We consider a double quantum dot system with an appreciable
tunnel coupling between the dots allowing for
coherent molecular states in these systems 
\cite{Oosterkamp}.
The computational basis is formed by the 
position states of an additional spin-polarized electron \cite{Hayashi,Hollenberg}. A superposition  
can be created by variation of the interdot coupling. 
In order to stabilize the charge, the coupling of the dots to the two leads is 
driven to weak values and the dot is tuned to the Coulomb
blockade regime \cite{Wilfred} where sequential 
tunneling is suppressed through the addition energy. 
Even then, the system couples to the environment through 
cotunneling \cite{Nazarov}, the correlated exchange of two electrons
with the external leads which ends up in a state with the same total
charge as the initial one.

The relevant Hilbert space is 
spanned by four states written 
$|i,j\rangle$ denoting $i,j$ additional
electrons on the left and right dot, respectively.
$|1,0\rangle$ and
$|0,1\rangle$ define the computational basis as they
are energetically accessible, the closest virtual 
intermediate states for cotunneling are
 $|v_0\rangle=|0,0\rangle$ and $|v_2\rangle=|1,1\rangle$. 
This model applies if all relevant energy scales of
the system ($\varepsilon_{\rm as}$ and $\gamma$, see below) are much smaller 
than the charging energies to the next virtual 
levels ($\varepsilon_{v_2}$ and $\varepsilon_{v_0}$, also below), which in 
turn have to be smaller than the orbital excitation of the individual dots. 
This can be realized in small dots.

\begin{figure}[htb]
\centering
\includegraphics[width=5truecm]{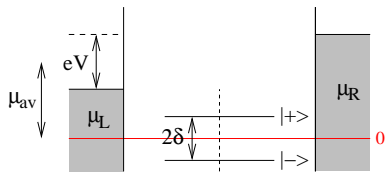}
\caption{(Color online) Sketch of the considered artificial molecule in the Coulomb Blockade 
regime, where $2\delta$ is the level splitting and $V=\mu_R-\mu_L$ the bias 
voltage, that is applied between the two leads (hatched). The virtual states
$|v_2\rangle$ and $|v_0\rangle$ are outside the plotted energy 
range.} 
\label{fig:molecule}
\end{figure}

The total Hamiltonian of this system can be written as
$H =H_{0} + H_{1}$ where $H_{0}= H_{{\rm sys}} + H_{{\rm res}}$ 
describes the energy spectrum of the isolated double-dot
through 
$H_{{\rm sys}} =  \varepsilon_{\rm as}(a^{L\dagger}a^L-a^{R\dagger}a^R)-
\varepsilon_{v_0} \hat{n}_{v_0}+\varepsilon_{v_2} \hat{n}_{v_2} 
+ \gamma (a^{L\dagger}a^R+a^{R\dagger}a^L)$
and the two electronic leads
$H_{{\rm res}}  =  \sum_{\vec{k}} \varepsilon^{L}_{\vec{k}}b^{L\dagger}_{\vec{k}}b^{L}_{\vec{k}} + \sum_{\vec{k\prime}} \varepsilon^{R}_{\vec{k\prime}}b^{R\dagger}_{\vec{k\prime}}b^{R}_{\vec{k\prime}}$. The sum over the dot states only 
runs over the restricted Hilbert space described above, 
the $a^{L/R}$ act on the lowest additional electron state on either dot. 
The double-dot is characterized by the asymmetry energy $\varepsilon_{\rm as}=\varepsilon_l-\varepsilon_r$ 
between the individual dots and the interdot tunnel coupling $\gamma$.
The virtual states $|v_2\rangle$ and $|v_0\rangle$ are separated from the  
system by energy differences $\varepsilon_{v_2}$ (upper virtual level) and 
$\varepsilon_{v_0}$ (lower virtual level).
The tunneling part $H_{1}=t_{c} \sum_{\vec{k},n} (a^{L\dagger}_{n}b^{L}_{\vec{k}}+a^{L}_{n}b^{L\dagger}_{\vec{k}}) 
+ t_{c} \sum_{\vec{k\prime},m} (a^{R\dagger}_{m}b^{R}_{\vec{k\prime}}+a^{R}_{m}b^{R\dagger}_{\vec{k\prime}})$ describes the
coupling of each dot to its lead and will be treated as a perturbation. 
$t_c$ represents the tunnel matrix elements between the dots and the leads. It can be absorbed in a tunneling rate $\hbar\Gamma=2\pi t_c^2 N(\epsilon_F)$. This has to be chosen small such that the
Kondo temperature is low $T_{\rm K}\ll T$ and perturbation theory holds, e.g.,
$\Gamma=10^9$~Hz. 
Figure~\ref{fig:molecule} shows a sketch of the setup.

From now on, we use the basis of molecular states
obtained by diagonalizing
$H_{\rm sys}$ with
splitting $2\delta=2\sqrt{\varepsilon_{\rm as}^2+\gamma^2}$. In order to 
treat cotunneling by leading-order perturbation theory, 
we rewrite $H_1$ using a Schrieffer-Wolff transformation
\cite{Schrieffer}. This removes the transitions to the virtual
states and 
generates an effective Hamiltonian containing indirect 
transition terms between the molecular
states. 
A more detailed description of our calculation is given 
in Refs.\ \onlinecite{UdoFrank} and \onlinecite{Diplom}. The final Hamiltonian
is of the form $H=H_0+H_1^\prime$ where
\begin{eqnarray}
H_1^\prime & = & \sum_{c,d}\limits \alpha_{c}^{\dagger}\alpha_{d} \Big[
\sum_{Y,Y',\vec{k},\vec{k'}}\limits H^{Y,Y'}_{\vec{k},\vec{k'},c,d}
b^{Y\dagger}_{\vec{k}} b^{Y'}_{\vec{k'}} + \nonumber \\
&& + \sum_{Y,Y',\vec{k},\vec{k'}}\limits
H^{Y,Y'}_{\vec{k},\vec{k'},c,d} b^{Y}_{\vec{k}} b^{Y'\dagger}_{\vec{k'}}\Big] ,
\label{eq:scatt_hamil}
\end{eqnarray} 
where $Y$ and $Y'$ denote right or left lead, the $\alpha$s 
describe molecular states and the 
$H^{Y,Y'}_{\vec{k},\vec{k'},c,d}$ are given through 
2nd order perturbation theory, i.e., they are of $O(\Gamma^2)$.
Note, that $H_1^\prime$ conserves the
particle number because it acts upon the double-dot by injecting {\em and}
extracting an electron in a single step. The terms with 
$Y\not=Y'$ transfer charge between different reservoirs. Note that
Eq.\ (\ref{eq:scatt_hamil}) is a simple and generic
 Hamiltonian connecting a quantum
system to two distinct particle reservoirs and is potentially relevant
for systems other than quantum dots as well. 

We study the open system dynamics in the case of a time-independent Hamiltonian
with a fully general initial reduced density matrix. 
We use the well-established and controlled Bloch-Redfield
\cite{Agyres}, which has been demonstrated to work down 
to low temperature for certain models \cite{Hartmann}.
It involves a Born approximation in $H^\prime_{\rm 1}$, i.e., it captures
all cotunneling processes in lowest nonvanishing order. 
The Redfield equations \cite{Weiss} for the elements of the reduced density 
matrix $\rho$ in the eigenstate basis of $H_{\rm sys}$ (i.e., the molecular 
basis) read
\begin{equation}
\dot{\rho}_{nm}(t)=-i\omega_{nm}\rho_{nm}(t)-\sum_{k,l}\limits R_{nmkl}\rho_{kl}(t) \ , 
\label{redfield}
\end{equation}
where $\omega_{nm}=(E_n-E_m)/\hbar$ and the Redfield tensor
elements $R_{nmkl}$ are composed of golden rule rates 
describing different cotunneling processes, 
which are essentially independent due to the low
symmetry of the system. Each contribution has a typical cotunneling 
structure \cite{UdoFrank,Diplom}. An overview of 
the most important processes is given below.
$n$, $m$, $k$ and $l$ can be either $+$
(excited molecular state) or $-$ (molecular ground state) with according 
energies $E_{\rm\pm}$. This type of perturbative analysis
is only valid above the Kondo temperature $T_K$\cite{Haldane}, which can be 
easily driven to low values by pinching off the tunneling barrier to the leads.

From the formal solution of Eq.\ (\ref{redfield})  
we can identify the relaxation and 
dephasing rates as
\begin{eqnarray}
\Gamma_{r} & = & \Re(R_{++++} + R_{----}) = \frac{1}{\tau_{r}} \label{eq:r}\\
\Gamma_{\phi} & = & \Re(R_{+-+-}) = \Re(R_{-+-+}) = \frac{1}{\tau_{\phi}} \label{eq:phi}\ .
\end{eqnarray}
The transition frequencies $\omega_{\rm nm}$ are 
weakly shifted.
\begin{figure}[h]
\begin{center}
\parbox{3.5cm}{ (a) \\
\includegraphics[width=3.5truecm]{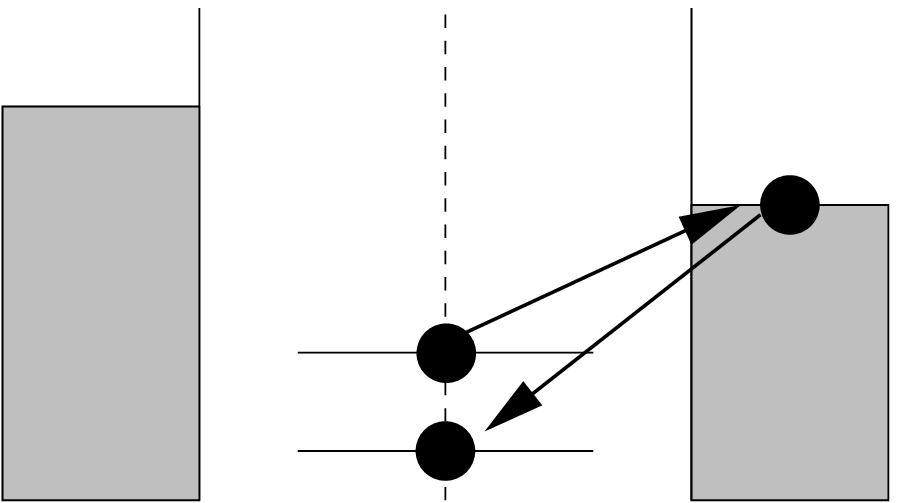}}
\hfill
\parbox{3.5cm}{ (b) \\
\includegraphics[width=3.5truecm]{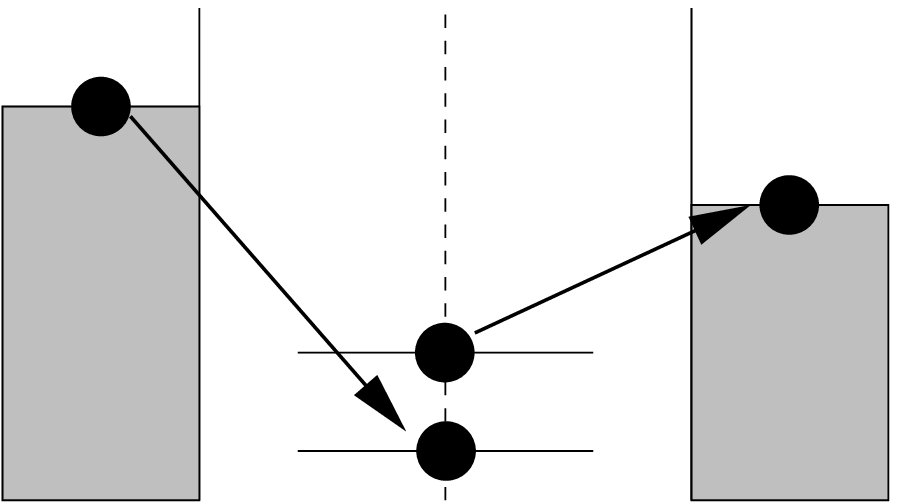}} \\
\vspace{0.4truecm}
\parbox{3.5cm}{ (c) \\
\includegraphics[width=3.5truecm]{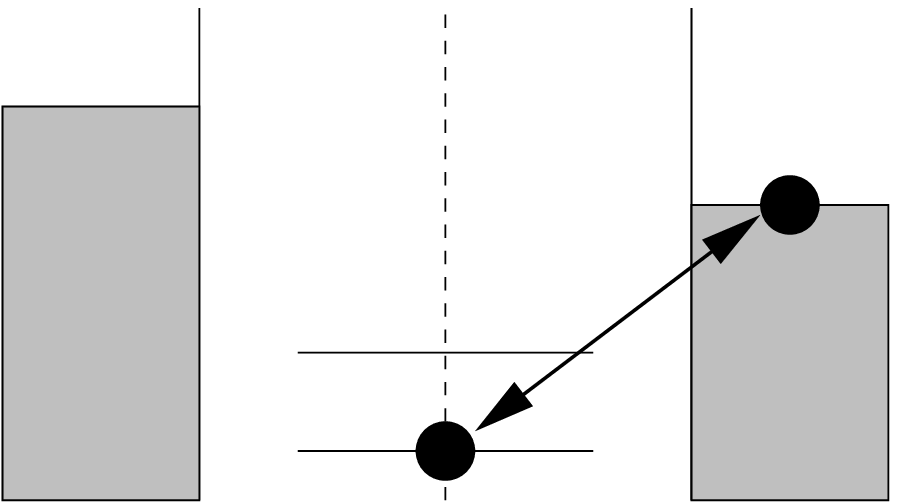}} 
\hfill
\parbox{3.5cm}{ (d) \\
\includegraphics[width=3.5truecm]{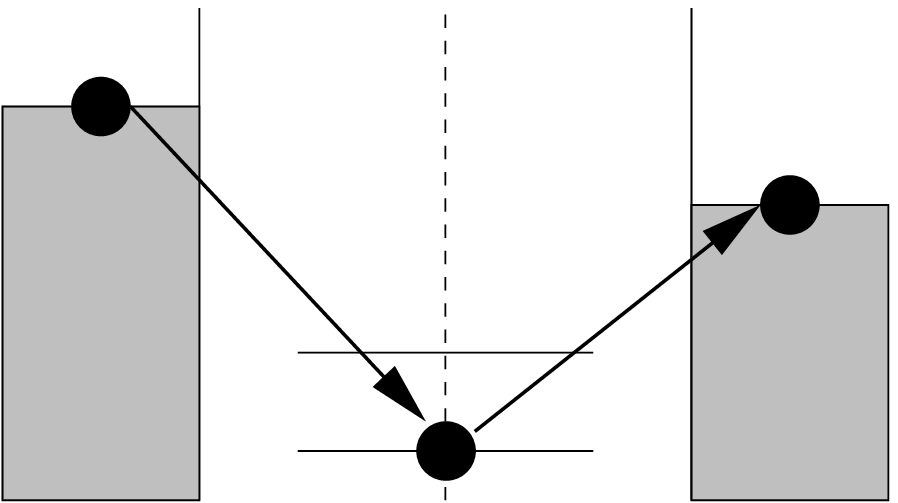}}
\end{center}
\caption{Examples for relevant processes in the system: (a) a relaxation 
process that carries no current, (b) a relaxation process with current, (c)
a pure dephasing process without current flow and (d) a current-carrying 
dephasing process.}
\label{fig:procs}
\end{figure}

Figure~\ref{fig:procs} shows a choice of processes
entering the Redfield tensor. 
All processes contribute to dephasing, because the phase of an
electron, which is injected from the reservoirs, is always random.
Figures (a) and (b) illustrate relaxation processes. 
Only (b) contributes to the 
 current, i.e., in general the {\em relaxation} rate must
not be confused with the cotunneling {\em current}. 
In (c) and (d) two pure dephasing processes are presented, only 
(d) contributes to the current flow. In general, processes without
current can emerge, if the cotunneling processes take place between a 
{\it single} lead and the two-state system (TSS). The {\em observable} current
is then given by the difference of current-carrying processes in forward and 
backward direction. 
We have evaluated the rates entering, 
Eqs.\ (\ref{eq:r})and (\ref{eq:phi}) using the 
$H^{Y,Y'}_{\vec{k},\vec{k'},c,d}$s. 
Due to the high number of terms, details are not shown and will be given 
elsewhere \cite{Diplom}.

\begin{figure}[h]
\centering
\includegraphics[width=\columnwidth]{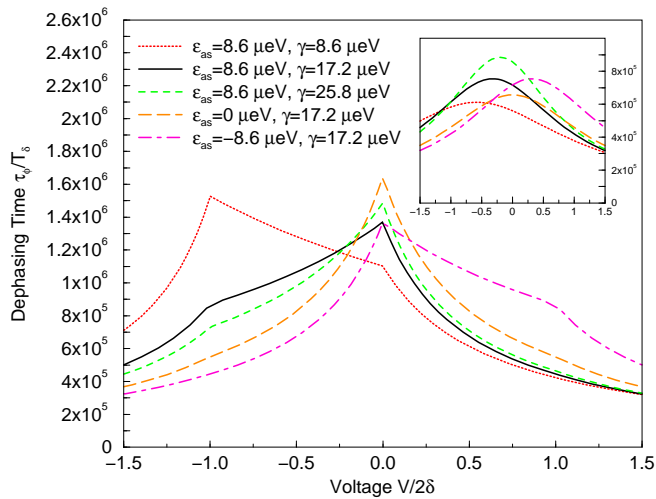}
\caption{(Color online) Dephasing time $\tau_{\phi}$ in the same units as in 
the previous figure for different values of 
$\varepsilon_{\rm as}$ and $\gamma$, when the bias voltage $V/2\delta$ is varied 
(with  $\mu_{\rm av}=(\mu_{R}+\mu_{L})/2=75.832$~$\mu$eV and $k_B T/\mu_{\rm av}=1.136\cdot10^{-3}$); inset: $k_B T/\mu_{\rm av}=0.159$.} 
\label{fig:deph_V}
\end{figure}

We now turn to the discussion of our results, starting with 
the relaxation time $\tau_{r}$.
\begin{figure}[h]
\centering
\includegraphics[width=\columnwidth]{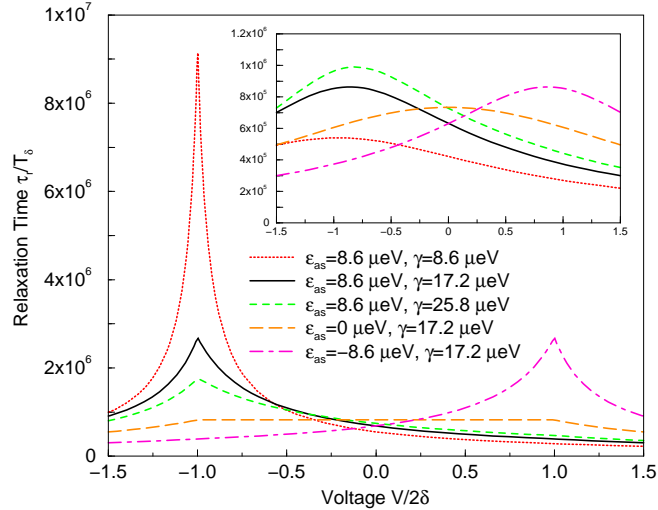}
\caption{(Color online) Relaxation time $\tau_{r}$ in 
units of $T_{\delta}=\frac{2\pi\hbar}{2\delta}$, 
the period of coherent oscillations between the two molecular
states. Different values of 
$\varepsilon_{\rm as}$ and $\gamma$ are taken, when the bias voltage $V/2\delta$ is varied 
(with  $\mu_{\rm av}=(\mu_{R}+\mu_{L})/2=75.832$~$\mu$eV and $k_B T/\mu_{\rm av}=1.136\cdot10^{-3}$); inset: $k_B T/\mu_{\rm av}=0.159$.} 
\label{fig:rel_V}
\end{figure}
We observe in Fig.~\ref{fig:rel_V} that for an asymmetric TSS, i.e., for 
$\varepsilon_{\rm as}\not=0$,
there is a pronounced peak of the relaxation time 
at $V=-{\rm sgn}(\varepsilon_{\rm as}) 2\delta$, where $V=\mu_R-\mu_L$,
i.e.\
the sign has to be chosen with opposite polarity to the asymmetry energy. 
This means in particular that the relaxation is minimal far away from 
equilibrium. This is the central result of our paper. It is most clearly 
visible for $T\ll 2\delta$, but obviously still dominates the calculated 
result for temperatures $T \approx 2\delta$, as it can be seen in the insets 
of Figures~\ref{fig:rel_V} and \ref{fig:deph_V}.
In order to remain in the cotunneling regime, the voltages are still quite 
small as 
compared to the excitation energy to the next charge states 
$\varepsilon_{v_2}$ and $\varepsilon_{v_0}$, but on the order of the molecular 
level splitting, i.e. 
$|V|\approx 2\delta\ll\varepsilon_{v_2},\varepsilon_{v_0}$. 
For quantum computation, achieving a maximum relaxation time is e.g.\
appreciable during {\em read-out} \cite{EPJ}. 

Although surprising, it can be understood from the 
analysis of the different rates, that $V=0$ does not
necessarily imply the lowest relaxation rate. At $V=0$
there is no {\em net} current, i.e.\ no {\em net} exchange of particles
in the ensemble average, however, this is achieved by the 
cancellation of finite currents of equal size in forward and backward
direction. 
These currents are rather small \cite{UdoFrank} such that current heating
is reduced to a minimum.
To $\tau_{\rm r}$, Eq. (\ref{eq:r}), such current-carrying processes
contribute with {\em equal} sign - the system relaxes no matter to which
reservoir. On top of this, one also has to take into account the
aforementioned current-less relaxation channels. 

\begin{figure}[h]
\begin{center}
\parbox{3.5cm}{ (I) \\
\includegraphics[width=3.5truecm]{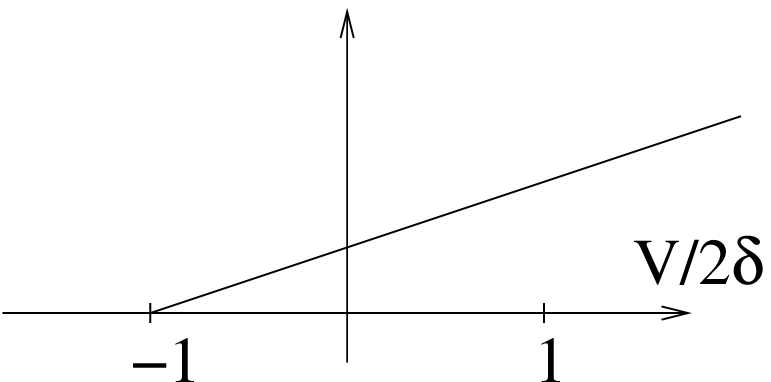}}
\hfill
\parbox{3.5cm}{ (II) \\
\includegraphics[width=3.5truecm]{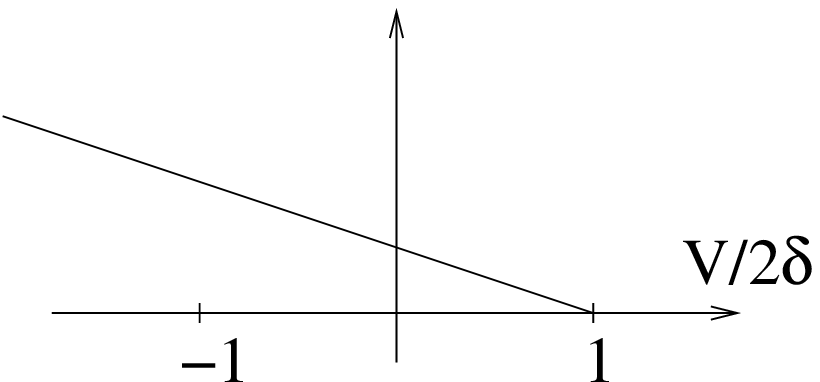}} \\
\vspace{0.4truecm}
\parbox{3.5cm}{ (III) \\
\includegraphics[width=3.5truecm]{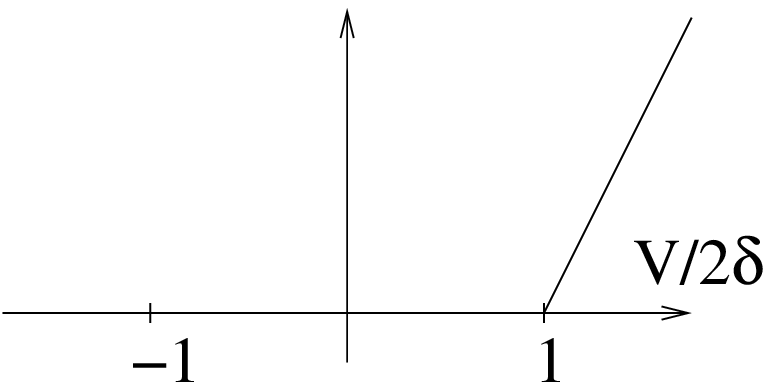}}
\hfill
\parbox{3.5cm}{ (IV) \\
\includegraphics[width=3.5truecm]{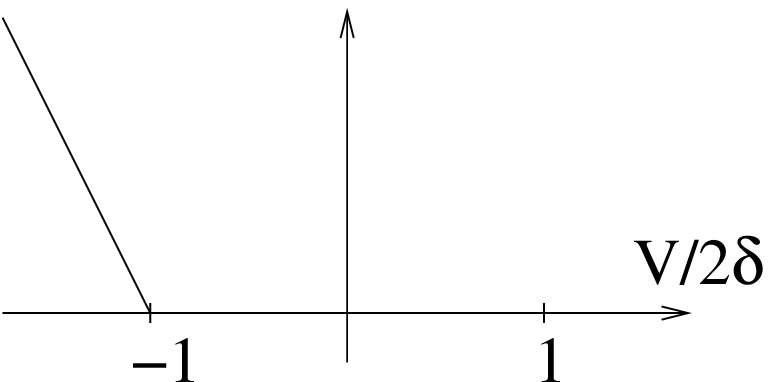}}
\end{center}
\caption{Qualitative voltage dependence of the rates of 
emission [(I), (II)] and absorption [(III), (IV)] processes, 
see text; these rates do not correspond to the processes in 
Fig.~\ref{fig:procs}}
\label{fig:rates}
\end{figure}

The appearance of the peaks as 
preferred stable points in Fig.~\ref{fig:rel_V} can be understood based
on the analysis of the current-carrying processes, 
[e.g. Figs.~\ref{fig:procs} (b) and (d)] as schematically shown in 
Fig.~\ref{fig:rates}. At low voltages, $|V|<2\delta$, 
the system relaxes into a thermal state close to the ground state. 
Relaxation takes place by 
spontaneous emission of energy into the environment and creation of an 
electron-hole pair in the leads. This pair can recombine through the 
electrical circuit which fixes the electrochemical potentials. This
leads to electrical current. 
As the voltage is increased away from 
$V=0$, emission processes which lead to a current 
{\em against} the polarity of the source are suppressed, the others are
enhanced, see Figs.~\ref{fig:rates} (I) and (II). Depending on the 
asymmetry of the double dot, i.e.\ on the weight of the excited state on the
left and the right dot, this leads to an enhancement or a suppression of 
the rate.  
At $|V|\ge2\delta$, the emission processes against the source 
are completely blocked: the dot relaxation does not provide enough energy
to overcome the electromotive force. The rate vanishes
{\em linearily} as a function of voltage reflecting the size of the
available phase space for cotunneling, see Fig.~\ref{fig:rates}. 

At higher voltages, $|V|\ge2\delta$, inelastic cotunneling \cite{Silvano}
sets in, see Figs.~\ref{fig:rates} (III) and (IV): 
The source provides enough energy to even excite the double dot, 
creating a nonequilibrium steady-state population of the molecular levels. 
Hence, inelastic
cotunneling provides a way for the dot to {\em absorb} energy from the
environment even at low temperature. This process
can be experimentally identified by a sharp increase of the current 
\cite{Silvano,UdoFrank}. 

Hence, at $V=\pm 2\delta$, three of the four processes depicted 
in Fig.~\ref{fig:rates} vanish at low temperatures, whereas at $V=0$ only
two vanish. The linear voltage-dependence of the rates leads to
the rather sharp cusps seen in Fig.~\ref{fig:rel_V}. This behavior
is smeared out at higher temperatures by thermal fluctuations. 
The peak height is set
by the remaining processes: Energy emission {\em with} the source and 
currentless relaxation, Fig.~\ref{fig:procs} (a). As explained above, 
the relative weight of the former strongly depends on the weight of the
excited molecular state on the individual dots and thus is responsible for
the strong asymmetry of the peaks in Fig.~\ref{fig:rel_V} for different 
asymmetry energies. 

Finally, we analyze the properties of the dephasing time $\tau_{\phi}$ as a 
function of the bias voltage.
The total dephasing rate contains relaxing as well as flipless (``elastic'')
processes. We hence observe in Fig.~\ref{fig:deph_V} a peak structure
at $V=-{\rm sgn}(\varepsilon_{as})2\delta$ as in the relaxation time, 
Fig.~\ref{fig:rel_V}, and a similar peak at $V=0$. The 
latter can be understood
from the suppression of flipless processes (energy exchange 0)in an
analogous way to the relaxation peak in Fig.~\ref{fig:rel_V} (energy exchange
$2\delta$).  
At low 
asymmetry energy $\epsilon_{\rm as}<\gamma$, 
the dephasing time at $V=0$ is longest. At high asymmetry 
$\epsilon_{\rm as}>\gamma$ and at the
nonequilibrium working point $V=-{\rm sgn} (\epsilon_{\rm as})2\delta$, 
$\tau_\phi$ is even longer.
In general, this indicates
the existence of two preferable working points for quantum computation:
One {\em in} equilibrium, the other again {\em far from} equilibrium. 
As also already seen in the inset of Fig.~\ref{fig:rel_V}, 
the voltage dependence at higher temperature is here smeared out
and the peaks merge.

A measurement of the relaxation and dephasing times should be feasible 
either by a time-resolved measurement of 
$\langle \sigma_{z}(t) \rangle$, e.g., through a single-electron transistor
or point contact \cite{Elzerman}, 
the saturation broadening method 
\cite{Marlies} or resonance schemes such as proposed in 
Ref.~[\onlinecite{Engel}] for spins.

Note that parts of the double-dot literature focus on decoherence through
phonons or photons (see Refs.\ \onlinecite{Qin,Oosterkamp,Hazelzet,Brandes}), 
whereas we focus on the cotunneling, which becomes relevant when phonons are
suppressed by a cavity \cite{Eva}.
If the {\em spin} in a dot is used
as qubit \cite{Elzerman,Loss}, cotunneling serves as an indirect contribution
to decoherence.

To conclude, we have studied the decoherence of charge states in 
a double quantum dot due to cotunneling.
We have shown that
decoherence can be controlled through a bias voltage $V$ (and thus creating a 
nonequilibrium situation) between the two fermionic baths. 
In particular, the optimum working point for read-out and potentially
also for operation of the qubit can be in an out-of-equilibrium situation
at a voltage $V=-{\rm sgn}(\varepsilon_{\rm as}) 2\delta$. We have given a 
consistent physical interpretation of our findings in terms of stability and
phase space. This effect of stabilization through nonequilebrium should
potentially be significant for other qubit candidates as well.

We thank J.~von Delft, L.~Borda, J.~K\"onig, M.J.~Storcz, A.W.~Holleitner,
A.K.~H\"uttel and E.M.~Weig for clarifying discussions. Work 
supported by SFB 631 of the DFG and in part by the National Security Agency 
(NSA) and Advanced Research and Development Activity (ARDA) under Army 
Research Office (ARO) contract number P-43385-PH-QC.


\end{document}